\documentstyle[12pt,twoside,fleqn,espcrc1]{article}

\newcommand{\AmS}{{\protect\the\textfont2
  A\kern-.1667em\lower.5ex\hbox{M}\kern-.125emS}}
\input epsf
\title{Monopole Current Dynamics and Color Confinement}

\author{H.~Ichie
\thanks{ichie@rcnp.osaka-u.ac.jp},
H.~Suganuma and A.~Tanaka
\address{Research Center for Nuclear Physics (RCNP), Osaka University\\
Mihogaoka 10-1, Ibaraki, Osaka 567, Japan}%
}

\begin{document}
% typeset front matter
\maketitle

\begin{abstract}
Color confinement can be understood 
by the dual Higgs theory, where monopole condensation leads to the exclusion 
of the electric flux from the QCD vacuum. 
We study
the role of the monopole for color confinement 
by investigating the monopole current system.
When the self-energy of the monopole current
is small enough,  long and complicated monopole 
world-lines appear, which is a signal of monopole condensation. 
In the dense monopole system, the Wilson loop obeys the  
area-law, and the string tension and the monopole density
have  similar behavior as the function of the self-energy,
which seems that monopole condensation 
leads to color confinement.
On the long-distance physics, the monopole current system almost
reproduces essential features of confinement properties in lattice QCD.
In the short-distance physics, however, 
the monopole-current theory would become nonlocal and complicated
due to the monopole size effect. This monopole size would provide a 
critical scale of QCD in terms of the dual Higgs mechanism.
\end{abstract}

\section{QCD-monopole for Quark Confinement}

Quantum Chromodynamics (QCD) is the fundamental theory of the strong 
interaction and describes not only hadrons but also their underlying 
structure in terms of quarks and gluons. 
In the high energy
region, the QCD gauge coupling becomes small, 
and therefore the system can be described by the perturbative QCD, 
where quarks and gluons are good degrees of freedom. 
On the other hand, in the low energy region, there are various 
nontrivial phenomena like color confinement, dynamical chiral-symmetry 
breaking and topological defects due to the strong coupling nature 
of QCD. However, the essential degrees of freedom are unclear in the 
nonperturbative QCD, and their finding is strongly desired for
understanding hadron physics.

In 1974, Nambu proposed an interesting idea that quark confinement 
can be interpreted using the dual version of the 
superconductivity\cite{nambu}.
In this picture, condensation of the color magnetic monopole is the key 
concept and leads to squeezing the color-electric flux between quarks 
through the dual Meissner effect. 
In 1981, 't~Hooft showed that QCD is reduced into an abelian gauge 
field theory with magnetic monopoles after the abelian gauge 
fixing\cite{thooft}. In this gauge, the only abelian gauge fields 
with monopoles would be essential for describing non-perturbative 
phenomena in the low energy region\cite{ezawa}-\cite{suganuma2}.

Recent lattice QCD simulations show  strong evidence on this dual 
Higgs picture for the nonperturbative QCD in the Maximally Abelian 
(MA) gauge\cite{diacomo}-\cite{suganuma2}. 
The MA gauge is considered as the best abelian gauge, and is realized 
by the gauge transformation so as to maximize the diagonal components 
of the gauge fields (or the link variables in the lattice formalism). 
In this gauge, information of the gauge configuration is 
concentrated into the diagonal components.
According to the nontrivial homotopy class, 
$\pi_2($SU$(N_c)/$U$(1)^{N_c-1})= Z^{N_c-1}_\infty$\cite{thooft}, 
the QCD-monopole appears from the hedgehog-like configuration, 
which accompanies large fluctuation of the gauge field in the QCD vacuum. 
Since the classical monopole mass is in inverse proportion to the 
gauge coupling constant, monopoles can be easily excited in the 
low energy region with the large gauge coupling.
Therefore,  QCD-monopoles  are expected to be the relevant degrees of 
freedom for the description of the non-perturbative QCD. 
However, such a  picture for the confinement mechanism in terms of 
monopole condensation can be applied only in the infrared region. 
In the ultraviolet region,
the large monopole mass leads to 
suppression of the monopole excitation, and the QCD system seems to be trivial
in terms of  topology.

In the dual Higgs phase, only short-range interactions remain 
among monopoles, and therefore the monopole current dynamics
is subjected to a simple action. 
Thus, the nonperturbative QCD would be described only by
the monopole degrees of freedom.
In this paper, we study the monopole current dynamics and color
confinement from the idealized monopole-current action.

\section{Monopole Current Dynamics and Kosterlitz-Thouless Type Transition}

We study the monopole current system using the idealized monopole action, 
and investigate the role of the
monopole currents for color confinement. 
In general, the 
current action includes nonlocal interaction terms. 
In the confinement phase, however, the nonlocal part of the interaction
can be neglected, 
because the currents are expected to interact each other only in the 
short distance due to the {\it screening effect} by the
dual Higgs mechanism\cite{ezawa}. 
With the current conservation, the partition functional is written as 
\begin{equation}
Z = \int {\it D}k_\mu {\rm exp} \{ -\alpha \int d^4x  {\rm tr}  k^2_\mu (x) \}
\delta(\partial_\mu k_\mu),
\label{eq:part}
\end{equation}
where 
$k_\mu(x) \equiv k^3_\mu(x)
\cdot \tau^3/2$ and
$\alpha$ are
the monopole current and
energy for the unit current length, respectively.
In order to perform the path integral (\ref{eq:part}),
we put the system on the 4-dimensional lattice with the lattice spacing 
%$a=\Lambda^{-1}$.
$a$.
In the lattice formalism, the monopole currents are defined on the dual 
lattice, $k^{\rm lat}_\mu (s) \equiv e /(4\pi) \cdot a^3 k^3_\mu(s)$.
The partition function is given as
%\begin{equation}
$
%Z = \sum_{k^{\rm lat}_\mu} e^{-\alpha^{\rm lat} 
Z = \sum_{k^{\rm lat}_\mu} {\rm exp} \{ -\alpha^{\rm lat} 
\sum_s  k_\mu^{\rm lat}(s) ^2 \} 
\delta (\partial_\mu k_\mu^{\rm lat}(s)),
$
%\end{equation}
where $\alpha^{\rm lat} \equiv \alpha / 2 \cdot (4\pi/ea)^2 $.

The lattice QCD simulation shows that 
one long monopole loop and many short loops appear in 
the confinement phase.
The only long loop becomes important 
for the properties of the vacuum,
while many short loops are originated 
from the fluctuation in the ultraviolet region and therefore 
can be neglected. 
In this system, the partition function can be approximated 
as the single monopole loop ensemble with the length $L$, 
%\begin{equation}
$
Z=\sum_{L}\rho(L)e^{-\alpha L},
$
%\end{equation}
where $L$ and $\rho(L)$ are length of the monopole loop and its 
configuration number, respectively. 
The monopole current with length $L$ is regarded as the $L$ step
self-avoiding random walk, where $2d-1=7$ direction is possible 
in each step.
Therefore, $\rho(L)$ is roughly estimated as $(2d-1)^L=7^L$.
Using this partition function, the expectation value of the monopole 
current length is found to be 
\begin{equation}  
\langle L \rangle =\frac{1}{Z}\sum_{L}\rho(L) L e^{-\alpha L} =
  \left\{ \begin{array}{ll}
  \{ \alpha-\ln(2d-1) \}^{-1}  & \mbox{if $\alpha > \ln(2d-1)$} \\
  \infty & \mbox{if $\alpha < \ln(2d-1)$}.
       \end{array}
  \right. 
\end{equation}
When the energy $\alpha$ is larger 
than the entropy $\ln(2d-1)$, the monopole loop length is finite. 
However, when $\alpha$ is smaller than the entropy, 
the monopole loop length 
becomes infinite, which is a signal of monopole condensation 
in the current representation.

In performing the simulation,
we construct the monopole current
as the sum of  plaquettes of the monopole current
because of the current conservation condition.
First, we prepare a random monopole 
current system (hot start) and no monopole current system (cold start). 
Then, we update the {\it link} of the monopole current using the Metropolis 
method.

We generate the monopole current system on $8^4$ lattices.  
Fig.1 shows the monopole current vacuum for the typical cases,
($\alpha = 1.5,1.8,1.9$) at a fixed time. 
Corresponding to each case, we show in Fig.2 the histograms of the 
monopole loop length\cite{bode} of one cluster with 40 configurations.
For large $\alpha$, only 
small loops  appear
and their density is small.
On the other hand, for small $\alpha$, 
there appear one long monopole loop and some short loops, and
the monopole currents are complicated and dense.
Between these two regions, there is critical value, $\alpha_c\simeq 
%{\rm ln}7
$  1.8,
where monopole loops with various length appear.
The monopole density 
$\rho_M \equiv \frac{1}{4V}\sum_{s,\mu} |k_\mu(s)|$ 
and the clustering parameter $\eta 
\equiv \frac{\sum_i L^2_i}{(\sum_i L_i)^2}$ 
are shown in Fig.3. and Fig.4, respectively.
Here, $L_i$ is the loop length of the $i$-th monopole  cluster.
The property of the clustering parameter is that it 
goes to unity when many loops cluster and unite to  one  
long loop, and it approaches to zero when  loops never cluster each other.
When  $\alpha$ becomes small, the monopole density  becomes
large gradually at $\alpha 
%\le
\stackrel{<}{\scriptstyle{\sim}}
 \alpha_c$.
However, the clustering parameter is drastically changed  at $\alpha_c$.
Quantitatively, the critical value of monopole current energy,
$\alpha_c \simeq 1.8$, is close to ${\rm ln7} \simeq 1.95$, which is
corresponding to 
 the entropy for the 1 step
 self-avoiding random walk. 
This vacuum can be roughly approximated with the vacuum made of the  longest 
monopole loop alone in terms of the mechanism of the transition.
Such a transition is quite similar to the 
Kosterlitz-Thouless type transition in 2-dimensional superconductors, where
vortex condensation may occur.

\section{Dual Field Formalism and Role of Monopoles for Confinement}
In this section, we study how these monopole currents 
contribute to the color confinement properties.
Quark confinement is characterized by the linear inter-quark potential,
which can be obtained from the area-law behavior of the Wilson 
loop,
%\cite{rothe}
$\langle W \rangle 
= \langle P {\rm exp}({ie\oint A_\mu dx_\mu}) \rangle$.
Therefore, it is desired to extract the gauge variable from the monopole 
current $k_\mu$.
We now derive the abelian gauge variable in stead of the non-abelian gauge 
field, because the lattice QCD results show
the color confinement phenomena can be discussed only with abelian part 
to some extent.
However,
in the presence of the magnetic monopoles, the ordinary abelian gauge 
field $A_\mu(x)$  inevitably includes the singularity as the Dirac string, 
which leads to some difficulties in the field theoretical treatment.
Instead, the dual field formalism is much useful to describe the monopole
current system, because the dual gauge field $B_\mu(x)$ can be introduced 
without the singularity for such a system with $k_\mu \neq 0$ and $j_\mu=0$.
This is the dual version of the ordinary gauge theory with $A_\mu(x)$
for the QED system with $j_\mu \neq 0$ and $k_\mu=0$.

Let us consider the derivation of the dual gauge field $B_\mu(x)$ from
the monopole current $k_\mu(x)$.
Using the relation
%\begin{equation}
$
\partial_{\mu}  {}^*F_{\mu\nu} = k_\nu
$
%\end{equation}
with ${}^*F_{\mu\nu} \equiv
\frac12 \epsilon_{\mu\nu\alpha\beta} F_{\alpha\beta}$,
%Without the electric currents, 
${}^*F_{\mu\nu}$ is given by
the 2-form of the dual gauge field $B_\mu$ as
%\begin{equation}
$
{}^*F_{\mu\nu} = \partial_\mu B_\nu - \partial_\nu B_\mu,
$
%\end{equation}
so that one finds
$
%\begin{equation}
\partial^2 B_\nu - \partial_\nu ( \partial_\mu B_\mu ) = k_\nu. 
%\end{equation}
$
By taking the dual Landau gauge, $\partial_\mu B_\mu=0$, 
this field equation becomes quite a simple form, 
$\partial^2 B_\mu = k_\mu$.
Therefore, starting from the monopole current configuration
$k_\mu(x)$,
we derive the dual gauge field $B_\mu$,
\begin{equation}
B_\mu (x) = \partial^{-2} k_\mu(x) = -\frac{1}{4\pi^2}
\int d^4 y \frac{k_\mu(y)}{|x-y|^2}. 
\label{eq:dual}
\end{equation}
Using the dual gauge field, the Wilson loop is obtained as
$
\langle W \rangle 
= \langle  {\rm exp} \{ {ie\oint A_\mu dx_\mu} \} \rangle 
%= \langle {\rm exp} 
%({ie\int F_{\mu\nu} d\sigma_{\mu\nu}}) \rangle \nonumber \\
 =  \langle {\rm exp} 
({ie \frac12 \epsilon_{\mu\nu\alpha\beta} \int d\sigma_{\mu\nu}
 {}^*F_{\alpha\beta} }) \rangle
 =  \langle {\rm exp} 
\{ {ie \frac12 \epsilon_{\mu\nu\alpha\beta} \int d\sigma_{\mu\nu}
(\partial_\alpha B_\beta - \partial_\beta B_\alpha) 
} \} \rangle.
$
%\end{eqnarray}

Now, we apply this dual field formalism to the monopole current 
system discussed in Section 2. 
Since the monopole current $k^{\rm lat}_\mu(s)$ is generated 
on the lattice with the mesh $a$, 
the continuous dual field $B_\mu(x)$ is derived from 
$k_\mu(x)\equiv 4\pi/e \cdot  k^{\rm lat}_\mu(s)\delta^3(x-s)$ 
using Eq.(\ref{eq:dual}), in principle. 
In estimating the integral in Eq.(\ref{eq:dual}) numerically, we use 
a fine lattice with a small mesh $c$.
{\it To extract} $B_\mu(x)$ {\it correctly, the mesh }$c$ {\it is to be taken enough 
small}.  However, too fine mesh is not necessary because 
the original current $k^{\rm lat}_\mu(s)$ includes the error 
in the order of $a$.
Numerical analyses show that the use of $c \simeq a$ is too 
crude for the correct estimation of the integral in Eq.(\ref{eq:dual}).
Instead, the calculation with $c \le a/2$ is good enough for the 
estimation of $B_\mu(x)$, and hence we take $c=a/2$ hereafter.
On lattices, the dual gauge field 
$\theta_\mu^{\rm dual} \equiv a e B_\mu /2$ 
is defined in the dual Landau gauge,
%\begin{equation}
$
\theta_\mu^{\rm dual} (s+\mu) \equiv 
2 \pi \sum_{s'}\Delta(s-s') k^{\rm lat}_\mu(s'),
$
using the lattice Coulomb propagator $\Delta(s)=(\partial_\mu
\partial'_\mu)^{-1}$, where $\partial'$ denotes the backward derivative.
The dual version of the abelian field 
strength $\theta_{\mu\nu}^{\rm dual} \equiv e a^2 {}^*F_{\mu\nu}/2$ 
is defined by
%\begin{equation}
$
\theta_{\mu\nu}^{\rm dual}(s) \equiv 
\partial'_\mu \theta^{\rm dual}_\nu(s) 
- \partial'_\nu \theta^{\rm dual}_\mu(s).
$
%\end{equation}

The expectation value of the Wilson loop $\langle W \rangle$ is shown in
Fig.5.
The Wilson loop exhibits the area-law and the linear confinement 
potential:
${\rm ln} \langle W \rangle$ decreases linearly with the inter-quark 
distance, where its slope corresponds to the string tension.
Quantitatively, the string tension is measured by the Creutz ratio,
%\begin{equation}
%$
%\chi(I,J) \equiv - {\rm ln} \{
%\frac{ \langle W(I,J) \rangle \langle W(I-1,J-1) \rangle }
%     { \langle W(I-1,J) \rangle \langle W(I,J-1) \rangle } \},
%$
%\end{equation}
and we show in Fig.6 $\chi(3,3)$ as a typical example.
For $\alpha < \alpha_c$, the string tension decreases as  $\alpha$
increases, and its behavior is similar to the lattice QCD result by setting 
$\alpha = 0.8 \beta$.
For $\alpha > \alpha_c$, the string tension becomes
almost vanishes.
% for $\alpha > \alpha_c$.
It is worth mentioning that the behavior of string tension or the Creutz ratio
becomes similarly to the monopole density shown in Fig.3.
Thus, the confinement force is controlled by the monopole density,
and therefore monopoles can be regarded as the essential degrees of
freedom for color confinement.

\section{Monopole Size and Critical Scale in QCD}

In the final section, we compare the monopole current system with  
abelian projected
QCD \cite{diacomo,poly} especially in terms of the coupling 
correspondence and the size of QCD-monopoles. 

To begin with, let us consider one magnetic monopole 
with an intrinsic radius $R$ in the multi-monopole system. 
In the static frame of the monopole, 
it creates a spherical magnetic field, 
%\begin{eqnarray}
%{\bf H}(r)  =
%  \left\{ \begin{array}{ll}
%
%  {g(r) \over 4\pi r^3} {\bf r}={{\bf r} \over e(r) r^3} 
% & \mbox{ for $ r \ge R $ } \\
% 
%  {g(r) \over 4\pi R^3} {\bf r}={{\bf r} \over e(r) R^3} 
% & \mbox{ for $ r \le R $ },
%       \end{array} 
%\right.
%\end{eqnarray}
$
{\bf H}(r)  =
  {g(r) \over 4\pi r^3} {\bf r}={{\bf r} \over e(r) r^3} 
$
for $ r \ge R $
and
$
{\bf H}(r)  =  {g(r) \over 4\pi R^3} {\bf r}={{\bf r} \over e(r) R^3} 
% & \mbox{ for $ r \le R $ }
$
for $ r \le R $,
where the QCD running gauge coupling $e(r)$ is used to 
include the vacuum polarization effect. 
Here, the effective magnetic-charge distribution is assumed to be 
constant inside the monopole for the simplicity. 

Now, we consider the lattice formalism with a large mesh $a>R$. 
The electromagnetic energy observed on the lattice around 
a monopole is roughly estimated as 
\begin{eqnarray}
M(a) \simeq \int_a^\infty d^3x {1 \over 2}{\bf H}(r)^2 
\simeq {g^2(a) \over 8\pi a}
={2\pi \over e^2(a)a},
\label{eq:massa}
\end{eqnarray}
which is largely changed depending on the lattice mesh $a$.
%
%which is finite owing to the lattice mesh $a$.
%
This simple estimation neglects the possible reduction of $g(r)$ 
in the infrared region due to the asymptotic freedom of QCD. 
The screening effect of the magnetic field by other monopoles 
also reduces $g(r)$ effectively in a dense monopole system. 
However, $M(a)$ is modified by at most factor 2 
($M(a) = \frac{\pi}{ e^2(a)a }$)
even for the screening case as $g(r)=g(a) \cdot \theta(2a-r)$. 
Then, even in the multi-monopole system, 
$M(a)$ would provide an approximate value for
the electromagnetic energy on lattices 
created by one monopole, 
and we call $M(a)$ as  `lattice monopole mass'.

For the large mesh $a>R$, the monopole contribution to the 
lattice action reads $S = M(a)a \cdot L$, 
where $L$ denotes the length of the monopole current 
measured in the lattice unit $a$. 
Therefore, $M(a)$ is closely related to the monopole-current coupling 
$\alpha^{\rm lat}$ 
%in
% Eq.(3)
%Eq.(\ref{eq:massa})
 and $\beta=2N_c/e^2$, 
%\begin{eqnarray}
$
\alpha^{\rm lat} \simeq M(a)a  \simeq {2\pi \over e^2(a)}
={\pi \over 2} \cdot \beta_{\rm SU(2)}.
$
%\end{eqnarray}
For the above screening case, this relation becomes 
$
\alpha^{\rm lat} \simeq M(a)a  \simeq {\pi \over e^2(a)}
={\pi \over 4} \cdot \beta_{\rm SU(2)},
$
which is consistent with the numerical result, $\alpha = 0.8\beta$,
discussed in Section 3.
Here, as long as the mesh is large as $a>R$, the lattice monopole 
action 
%Eq.(\ref{eq:part})
would not need modification by the monopole size effect, 
and the current coupling $\alpha^{\rm lat}$ is proportional to $\beta$.
%
%
%
%In terms of the renormalization group, 
%we consider correspondence between the QCD gauge coupling and 
%the current coupling $\alpha^{\rm lat}$ 
%by regarding the change of $a$ as the scale transformation.
%First, we discuss the infrared (strong coupling) region, 
%
%at a large distance scale with a large scale unit $a$, 
%
%where the scale unit $a$ can be taken large as $a¨‡$. 
%In the lattice QCD, $\beta$ decreases due to the asymptotic freedom, 
%and hence the lattice-QCD action becomes 
%less important than $\Pi_{s,\mu} dU_\mu(s)$,
%
%the configuration number on the link variable $U_\mu(s)$, 
%
%which leads to the random color system 
%as the result of the `entropy dominance' 
%on the link variable $U_\mu(s)$. 
%
%Similar situation can be found in the lattice current dynamics: 
%since $\alpha^{\rm lat}$ decreases like $\beta$, 
%the lattice current action becomes less important 
%than $\Pi_{s,\mu} dk_\mu(s) \delta(\partial_\mu k_\mu)$, 
%
%the configuration number on the current variable $k_\mu(s)$, 
%
%which leads to random monopole-current generation 
%due to the `entropy dominance' on the current variable $k_\mu(s)$.
%
%In terms of the monopole mass  $M(a)$ in Eq.(\ref{eq:massa}), 
%the QCD-monopole becomes almost massless in the infrared region, 
%so that such a topological excitation can be easily done 
%and becomes important. 
Quantum mechanically, 
there is the energy fluctuation about $a^{-1}$ at the scale $a$, 
and therefore monopole excitation occurs very often at the long-distance 
scale satisfying $M(a) 
%\lsim
\stackrel{<}{\scriptstyle{\sim}}
 a^{-1}$. 
Thus, $M(a)a  \simeq \alpha^{\rm lat} 
%\simeq 2\pi/e^2(a)
$ 
is the control parameter for the monopole excitation 
at the scale $a(>R)$, 
and we can obtain the quantitative criterion 
for `monopole condensation' as $M(a)a \stackrel{<}{\scriptstyle{\sim}}
 {\rm ln}(2d-1)$ 
from the analysis using the current dynamics in Section 2. 

Second, we discuss the ultraviolet region with $a < R$.
%by moving the scale unit $a$ as $a
%\stackrel{<}{\scriptstyle{\sim}} R
%In the lattice QCD, $\beta$ increases as $\alpha$, 
%so that the lattice-QCD action becomes the key factor instead of 
%$\Pi_{s,\mu}dU_\mu(s)$.
%In the lattice current dynamics, 
%as long as $a>R$, $\alpha^{\rm lat}$ also increases like $\beta$, 
%and the lattice current action becomes more crucial than 
%the current configuration number. 
%Since the monopole mass $M(a)$ increases faster than the energy 
%fluctuation $a^{-1}$ as $a
%\stackrel{<}{\scriptstyle{\sim}} R$, monopole excitation is relatively 
%suppressed and becomes irrelevant. 
%
In the current dynamics, 
%however, 
there exists a critical coupling $\alpha_c \simeq {\rm ln} (2d-1)$ 
as shown in Section 2 and 3. 
Above $\alpha_c$, the lattice current action 
%in (\ref{eq:part}) 
provides 
no monopole condensation and no confinement, while 
$\beta \rightarrow \infty $ can be taken in the original QCD 
keeping the confinement property shown in Fig. 6. 
%
%the lattice unit $a$ becomes zero for $ƒ¿ \ge ƒ¿_c$.
%
Such a discrepancy between $\beta$ and $\alpha$ 
can be naturally interpreted by introducing the monopole size effect.
Obviously, the monopole-current theory should be drastically 
changed in the ultraviolet region as $a<R$, 
if the QCD-monopole has its peculiar size $R$. 

%
%Paying attention to the `determination' of the scale unit $a$, 
%
Here, let us reconsider the relation between $a$ and 
$\alpha^{\rm lat}$ in the lattice current action. 
%Eq.(\ref{eq:massa}). 
Similarly in the lattice QCD, the action 
%Eq.(\ref{eq:massa})
 has no definite 
scale except for the lattice mesh $a (>R)$, and therefore 
the scale unit $a$ would be determined so as to reproduce 
a suitable dimensional variable, 
e.g. the string tension $\sigma \simeq$ 1GeV/fm, 
in the monopole current dynamics. 
For instance, $a$ is determined as a function of $\alpha^{\rm lat}$ 
using the Creutz ratio $\chi \simeq \sigma a^2$ in Fig.6. 
Therefore, $a$ should reach $R$ before realizing 
$\alpha^{\rm lat} \rightarrow \alpha^{\rm lat}_c$, 
and the framework of the current theory is to be largely 
modified due to the monopole size effect for $a<R$.

Finally, we consider the fine lattice with a small mesh $a<R$, 
and study the monopole size effect to the current action 
(\ref{eq:part}).
In this case, the extended structure of the QCD-monopole is observed 
on the lattice, and the monopole creates the electromagnetic energy 
\begin{eqnarray}
M(a) \simeq \int_a^\infty d^3x {1 \over 2}{\bf H}(r)^2 
\simeq {g^2(R) \over 8\pi R}+ \int_a^R drr^4 {g^2(r) \over 8\pi R^6}
%\simeq {6 \over 5}{g^2(R) \over 8\pi R}
%      ={6 \over 5}{2\pi \over e^2(R) R},
\simeq {g^2(R) \over 8\pi R}={2\pi \over e^2(R) R} 
\end{eqnarray}
on the lattice.
Then, the `lattice monopole mass' $M(a)$ is determined mainly 
by the size $R$, and is almost $a$-independent. 
Accordingly, the lattice action should be changed as 
\begin{eqnarray}
S=M(R) a \sum_{(s,s'),\mu} k_\mu(s)k_\mu(s') \cdot
\theta \left( {R \over a}-|s-s'| \right) \delta_{s_\mu, s'_\mu} 
\end{eqnarray}
because of the self-avoidness originating from the monopole size $R$.
Thus, the monopole-current theory becomes nonlocal 
in the ultraviolet region $a<R$.

In conclusion, the QCD-monopole size $R$ provides a critical scale 
for the description of QCD in terms of the dual Higgs mechanism. 
In the infrared region as $a>R$, 
QCD can be approximated as a local monopole-current action\cite{ezawa}, 
and the QCD vacuum can be regarded as the dual superconductor. 
On the other hand, in the ultraviolet region as $a<R$, 
the monopole theory becomes nonlocal and complicated 
due to the monopole size effect, and the perturbative QCD 
would be applicable instead.

We acknowledge Prof. H.~Toki, Dr. A.~Hosaka and Prof.~Z.~Ma for 
their kind encouragement. One of authors (H.I.) is supported by 
Research Fellowships of the Japan Society for the 
Promotion of Science for Young Scientists.

%\newpage

\noindent {\bf \large Figure Captions} 

\noindent {\bf Fig.1} 
The monopole currents for  typical cases, (a) $\alpha=1.9$,
(b) $\alpha=1.8$, (c) $\alpha=1.5$. 

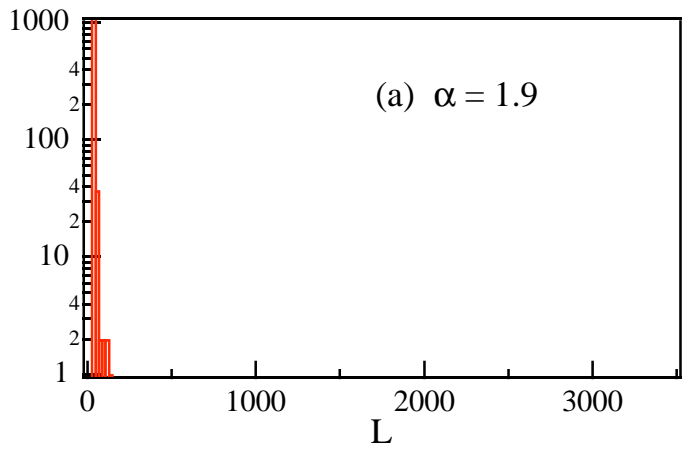
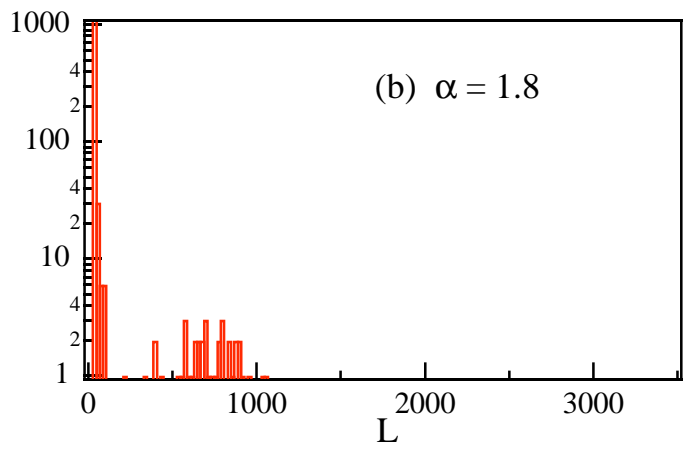
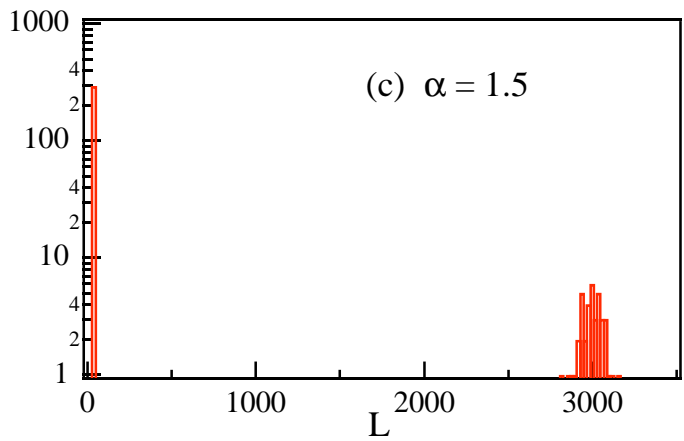
\noindent {\bf Fig.2} 
The histograms of the monopole current length of each cluster
with 40 configurations for the same cases with Fig.1. 

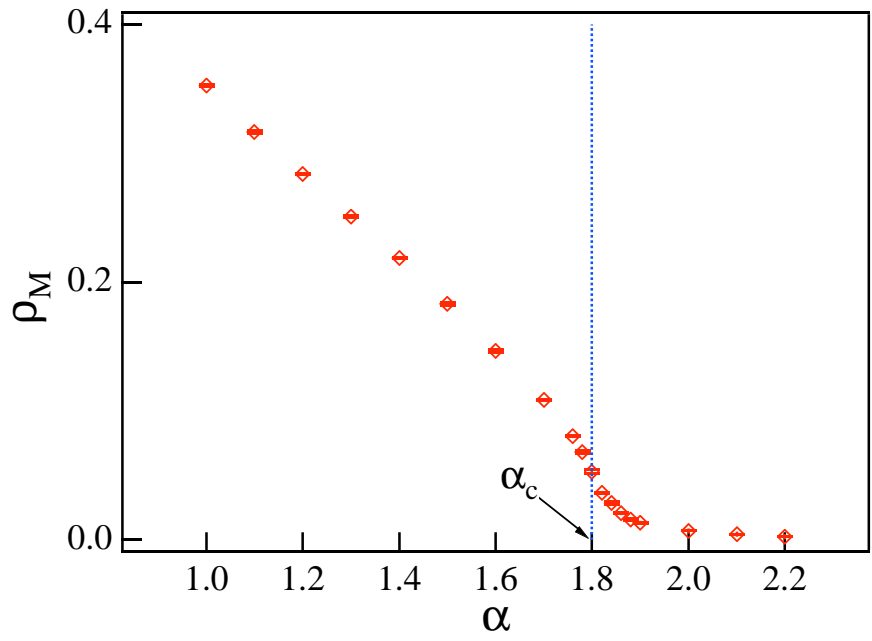
\noindent {\bf Fig.3}
The monopole density $\rho_M$  as the function of $\alpha$.
The monopole density becomes 
small for large $\alpha$ and almost vanishes for $\alpha > \alpha_c$. 

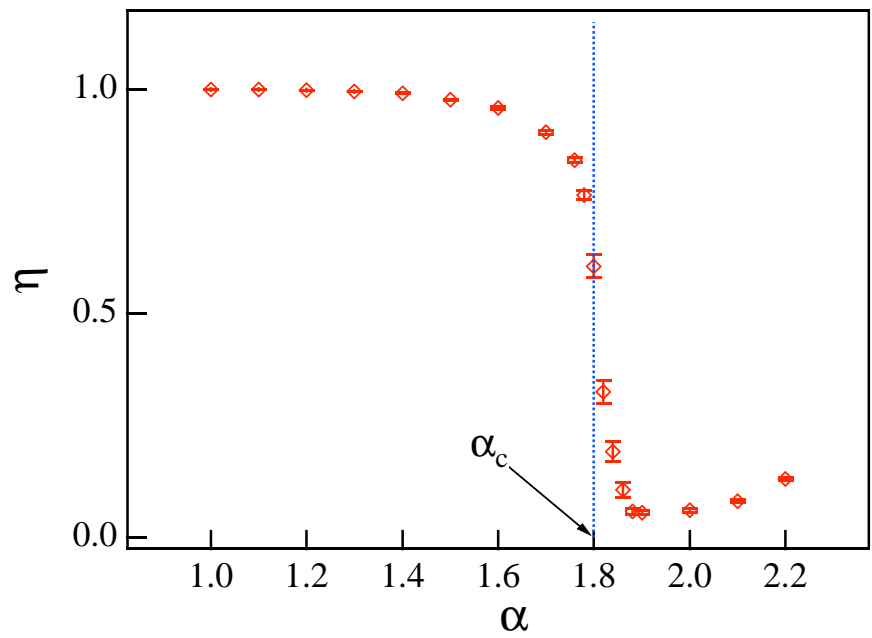
\noindent {\bf Fig.4}
The clustering parameter $\eta$ as the function of $\alpha$.
At $\alpha_c$, $\eta$ is drastically changed. 

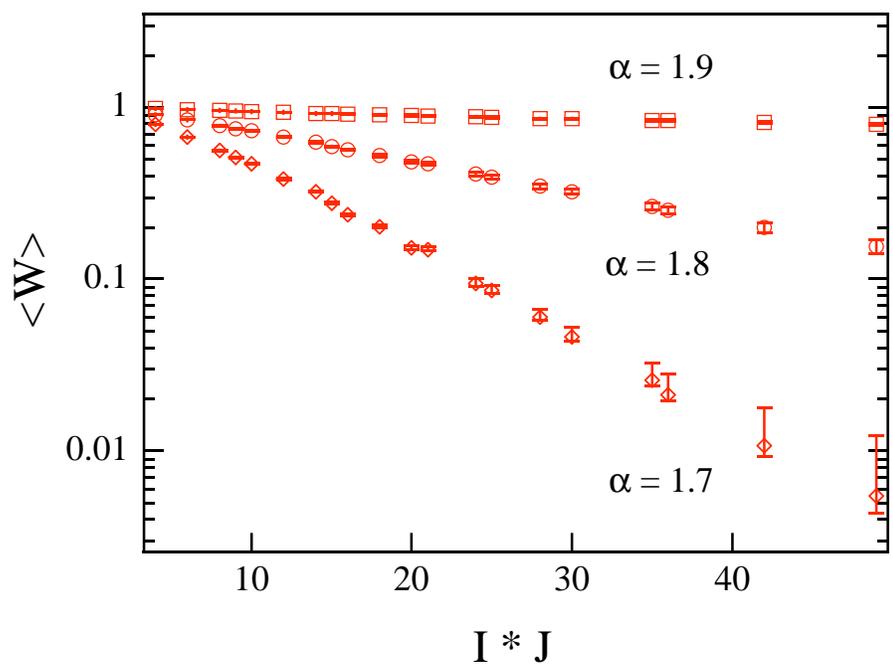
\noindent {\bf Fig.5}
The expectation value of the Wilson loop 
$\langle W(I \times J) \rangle$ on fine mesh $c=a/2$ for $\alpha=1.7,1.8,1.9$.
The area-law behavior indicates the linear interquark potential. 

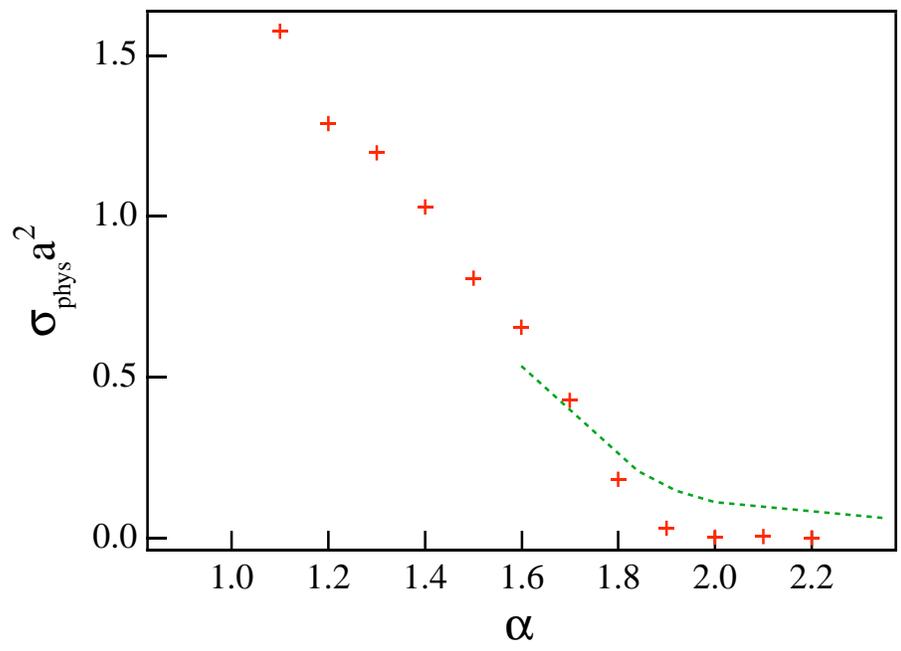
\noindent {\bf Fig.6}
The Creutz ratio as the function of $\alpha$ in the multi-monopole system.
The dotted line denotes the Creutz ratio as the function of $\alpha = 
0.8\beta$ in the lattice QCD.


\begin{thebibliography}{9}

     \bibitem{nambu}
Y.~Nambu, Phys.~Rev.~{\bf D10} (1974) 4262.
     \bibitem{thooft}
G.~'t~Hooft, Nucl.~Phys.~{\bf B190} (1981) 455.
     \bibitem{ezawa}
Z.~F.~Ezawa and A.~Iwazaki,
Phys.~Rev.~{\bf D25} (1982) 2681 ; {\bf D26} (1982) 631.
     \bibitem{suganuma}
H.~Suganuma, H.~Toki, S.~Sasaki,
% \&
H.~Ichie, Prog.~Theor.~Phys.
(Suppl.) {\bf 120} ('95) 57. \\
H.~Suganuma, S.~Sasaki and H.~Toki,
Nucl.~Phys.~{\bf B435} (1995) 207.
     \bibitem{ichiea}
H.~Ichie, H.~Suganuma and H.~Toki, Phys.~Rev. {\bf D52} (1995) 2994; 
{\bf D54} (1996) 3382. 
     \bibitem{diacomo}
A.~Di~Giacomo,  Nucl.~Phys.~{\bf B} (Proc.Suppl.) {\bf 47} (1996) 136
and references therein.    
     \bibitem{poly}
M.~I.~Polikarpov, Nucl.~Phys.~{\bf B} (Proc.Suppl.) {\bf 53} (1997) 134
and references therein.
      \bibitem{miyamura}
O.~Miyamura, Phys.~Lett.~{\bf B353} (1995) 91;
Nucl.~Phys.~{\bf B} (Proc.Suppl.) {\bf 42} ('95) 538.
      \bibitem{suganuma2}
H.Suganuma, A.Tanaka, S.Sasaki and O.Miyamura, 
Nucl. Phys.{\bf B} (PS) {\bf 47} ('96) 302.
      \bibitem{bode}
A.~Bode, T.~Lipper and K.~Schilling, 
Nucl.~Phys.~{\bf B} (Proc.~Suppl.) {\bf 34} (1994) 549.     
      \bibitem{ichie}
H.~Ichie and H.~Suganuma, Proc. of INSAM Symp. '96, Hiroshima, 
INSAM report.



\end{thebibliography}
\end{document}